\documentclass [11pt] {article}

\usepackage {times}
\usepackage [pdftex] {graphicx}
\usepackage {hyperref}

\begin {document}

\begin {center}
{\huge Stochastic Characteristics and Simulation of the Random Waypoint Mobility Model } \\
\vspace*{10pt}
{\large Ahuja, Aditya $^{1}$, Venkateswarlu K. $^{1}$, and Venkata Krishna, P. $^{1}$} \\
{\large $^{1}$School of Computing Science and Engineering, VIT University, Vellore - 632 014 } \\
{\large aditya.ahuja@intel.com, venkateswarlu.vit@gmail.com, parimalavk@gmail.com} \\
\vspace*{10pt}
\end {center}

\section* {Abstract}
Simulation results for Mobile Ad-Hoc Networks (MANETs) are fundamentally governed by the underlying Mobility Model. Thus it is imperative to find whether events functionally dependent on the mobility model `converge' to well defined functions or constants. This shall ensure the long-run consistency among simulation performed by disparate parties. This paper reviews a work on the discrete Random Waypoint Mobility Model (RWMM), addressing its long run stochastic stability. It is proved that each model in the targeted discrete class of the RWMM satisfies Birkhoff's pointwise ergodic theorem \cite{WikiErg}, and hence time averaged functions on the mobility model surely converge. We also simulate the most common and general version of the RWMM to give insight into its working.

\textbf{Keywords:} Random Waypoint Mobility Model, Asymptotic Mean Stationary, Ergodic, Simulation

\section* {Introduction}
Mobility models are used for the generation of node movement in simulations of MANETs. Protocol development is a consequence of such a simulation. The probabilistic aspects of the founding mobility model has direct implications on the simulation results. Many papers \cite{Ref2}-\cite{Ref5} have already concluded that stochastically unstable mobility models shall result in simulation results that diverge in time. 

The Random Trip Mobility Model, through the presence of a unique stationary distribution for the location of nodes, has already been proved to be stable \cite{Ref5}. \emph{The work presented in this paper is purely a review of the stability of the discrete version of the RWMM proved by Timo, Blackmore and Hanlen} \cite{Main}. Therein the notion of stability is considered to be the satisfaction of Birkhoff's Pointwise Ergodic Theorem by the mobility model. If to the contrary the mobility model is unstable, the simulation results are bound to be unreliable.

The stimulus for this line of work is that the stability or lack thereof of the mobility model is possibly passed up the layers of the protocol stack. For instance the DSR protocol preserves the mobility model's stability \cite{Ref6}: if the node location random process is stable, then so is the route selection random process. The consequence of this is that the strong law of large numbers also holds for the simulations at the network layer.

 A mobility model is quantified using a random process. It is stationary if the set of probability laws regulating the movement of the nodes are independent of time. Many works have come up with the transformation of non-stationary models into (in some places pointwise ergodic theorem satisfying) stationary models with the motivation that the strong law of large numbers may be applicable.

The classic RWM model does display starting transients and local nonstationarity. Thus we analyze its properties by means of imposing a mathematically weaker `asymptotic stationarity' property. A random process, the mean of which is governed asymptotically by a process with a stationary distribution is called Asymptotically Mean Stationary (AMS). It has been proved \cite{Ref9}[Theorem 1] that a random process is AMS if and only if it satisfies Birkhoff's pointwise ergodic theorem. By consequence a mobility model is stable if and only if it is AMS \cite{Main}.

In the classic RWMM \cite{Main}, every node, using an independent and identically uniformly distributed (IID) random process $\{W_k\}_{k=0}^\infty$, selects a sequence of waypoints $ \mathbf{w} = w_0,w_1,w_2,... $. For every pair $(w_i, w_{i+1}), i \in \mathcal{Z^*}$, the node chooses a speed randomly and uniformly distributed from the closed interval $[min\_s, max\_s]$. At this chosen speed it then travels in a straight line from $w_i$ to $w_{i+1}$.

In this review, the main result addressed is: a) The general discrete class of the RWMM is asymptotically mean stationary (by virtue of which it is stable) and ergodic. b) For stable node movement the following conditions suffice - (i) Node waypoint selection is an AMS random process, (ii) Speed selection random process is stationary.

This paper is organized as follows. The next section introduces the preliminaries. Following that we describe the general RWMM. Next is the contribution of the base paper in the form of a theorem. Simulation results for the classic case and conclusion end this paper. \\

\section* {Preliminaries}
We will adopt the dynamical systems \cite{Ref11}-\cite{Ref12} model for a random process. Given a discrete finite alphabet $\mathcal{X}$, let $\mathbf{X} = \{X_k\}_{k=0}^\infty$ be the associated discrete time random process. The distribution of  $\{X_n\}_{n=0}^\infty$ is the set $\{\mu^{(k)} : k \geq 0 \}$ where $\mu^{(k)}$ is the probability measure on $\mathcal{X}^k$ given by: \\ $\mu^{(k)}(x_0^{k-1}) = Pr [ X_0 = x_0, X_1 = x_1, ..., X_{k-1} = x_{k-1} ]$ 

In order to simplify our work, we use the Kolmogorov Representation Theorem (where certain consistency conditions are satisfied) \cite{Ref11}[Theorem I.1.2]. This enables us to replace the distribution with a unique probability measure $\mu$ on the  space $\mathcal{X}^\infty = \Pi_{i=0}^\infty \mathcal{X}$. Throughout we shall be dealing with cylinder sets as elementary events:  $ [x_m^n] = \{\mathbf{\bar{x}} : \bar{x_i} = x_i, m \le i \le n \} $. The $\sigma$-algebra $\mathcal{F_{X^\infty}}$ is generated using these cylinder sets. Time is incorporated using the shift transform 
$ T_\mathcal{X}^k = x_k, x_{k+1}, x_{k+2}, ... , k \in \mathcal{Z^*} $. Eventually we result with the dynamical system 
$ (\mathcal{X^{\infty}}, \mathcal{F_{X^\infty}}, \mu, T_{\mathcal{X}}) $ which is related to the original random process by 
$ \{X_k\}_{k=0}^\infty = \{\Pi_0(T_\mathcal{X}^k\mathbf{x})\}_{k=0}^\infty $, $\Pi_0\mathbf{x} = x_0 $. 

Suppose we have a mobility model quantified by the random process $\{X_k\}_{k=0}^\infty$ and  capture the location of each node for the first $k$ time instances of a simulation given by $x_0^{k-1} = x_0,x_1,x_2,...,x_{k-1}$. The dynamical system associated with this stochastic experiment is
$ (\mathcal{X^{\infty}}, \mathcal{F_{X^{\infty}}}, \mu, T_\mathcal{X})$ and the trajectory captured is the elementary event $ [x_0^{k-1}] \in \mathcal{F_{X^\infty}} $. If variable length shift must be considered 
$ T_\mathcal{X^*}\mathbf{x} =  T_\mathcal{X}^{f\mathbf{(x)}}\mathbf{x} $ as is necessitated in certain cases by the random processes associated with the updation of routing tables of network routers, we may study the probabilistic properties of $ (\mathcal{X^{\infty}}, \mathcal{F_{X^\infty}}, \mu, T_{\mathcal{X^*}}) $.

Now we come up with certain definitions and lemmas lifted from the base work which serve as the foundation for future proof developments. \\
\emph{Definition 1 (Stationarity)}\cite{Main}: The system $ (\mathcal{X^{\infty}}, \mathcal{F_{X^{\infty}}}, \mu, T_\mathcal{X})$ is called stationary, and $T_\mathcal{X}$ is said to be measure preserving if, $ \forall A \in \mathcal{F_{X^\infty}} $,  $ \mu(A) = \mu(T^{-1}A) $ .\\
\emph{Definition 2 (Ergodicity)}\cite{Main}\cite{WikiErg}: The stationary system $  (\mathcal{X^{\infty}}, \mathcal{F_{X^{\infty}}}, \mu, T_{\mathcal{X}}) $ is said to be ergodic if   $ A = T^{-1}A \Rightarrow \mu(A) = 0 $ or $1$ . Equivalently, 
$ (\mathcal{X^{\infty}}, \mathcal{F_{X^{\infty}}}, \mu, T_{\mathcal{X}}) $ is ergodic iff $ \forall f \in L^1(\mu) $ the limit \\
$<f> = <f>(\mathbf{x}) = \lim\limits_{n \to \infty} \frac{1}{n} \sum\limits_{k=0}^{n-1} f(T^k_\mathcal{X}\mathbf{x})$ 
\hspace*{10pt} is a constant almost everywhere in $\mu$. \\
\emph{Definition 3 (Stability)}\cite{Main}: A mobility model associated with the random process 
 $ (\mathcal{X^{\infty}}, \mathcal{F_{X^{\infty}}}, \mu, T_{\mathcal{X}}) $ is said to be stable if for all bounded and measurable $f$, the limit \\
$<f>(\mathbf{x}) = \lim\limits_{n \to \infty} \frac{1}{n} \sum\limits_{k=0}^{n-1} f(T^k_\mathcal{X}\mathbf{x})$
\hspace*{10pt} exists almost everywhere in $\mu$. \\
\emph{Definition 4 (Asymptotic Mean Stationarity)}\cite{Main}: The system $  (\mathcal{X^\infty}, \mathcal{F_{X^\infty}}, \mu, T_\mathcal{X}) $ is said to be asymptotic mean stationary (AMS) if, $ \forall A \in \mathcal{F_{X^\infty}} $ the limit \hspace*{10pt}$ \overline{\mu}(A) = \lim\limits_{n \to \infty} \frac{1}{n} \sum\limits_{k=0}^{n-1}\mu(T^{-k}_\mathcal{X}A) $  \hspace*{10pt} exists.\\ Here the probability measure $ \overline{\mu} $ is defined on the measurable space $ (\mathcal{X^\infty}, \mathcal{F_{X^\infty}}) $. It is called the stationary mean of $\mu$ and describes the average of the long run behaviour of the system. \\
\emph{Lemma 1 (Birkhoff's Pointwise Ergodic Theorem) } \cite{Main}\cite{WikiErg}: Let the dynamical system $ (\mathcal{X^\infty}, \mathcal{F_{X^\infty}}, \mu, T_\mathcal{X}) $ have $ T_\mathcal{X} $ as a measure preserving map, and let $f$ be measurable with $E(|f|) < +\infty$. Then \\ $\lim\limits_{n \to \infty} \frac{1}{n} \sum\limits_{k=0}^{n-1} f(T^k_\mathcal{X}\mathbf{x}) = E(f|\mathcal{C})$. \hspace*{5pt} Here $\mathcal{C}$ is the $\sigma$-algebra of invariant sets of $T_\mathcal{X}$ . If the random process is ergodic, then $\mathcal{C}$ is the trivial $\sigma$-algebra, and 
$  E(f|\mathcal{C}) =  E(f) $ which is a constant. \\
It has been proved \cite{Ref9} that asymptotic mean stationarity is both a necessary and sufficient condition for the pointwise ergodic theorem. \\
\emph { Lemma 2 (AMS Pointwise Ergodic Theorem) } \cite{Ref9}[Theorem 1]: A dynamical system  $ (\mathcal{X^\infty}, \mathcal{F_{X^\infty}}, \mu, T_\mathcal{X}) $ is AMS iff for all measurable $f$ with a finite expectation, the limit \\
$<f>(\mathbf{x}) = \lim\limits_{n \to \infty} \frac{1}{n} \sum\limits_{k=0}^{n-1} f(T^k_\mathcal{X}\mathbf{x})$
\hspace*{10pt} exists almost everywhere in $\mu$. \\
Eventually we conclude, using definitions 3,4 and lemma 2: \\
\textsc{Stability:} A mobility model with $  (\mathcal{X^{\infty}}, \mathcal{F_{X^{\infty}}}, \mu) $ as the  associated probability space is stable with respect to $T_{\mathcal{X}}$ iff   $  (\mathcal{X^{\infty}}, \mathcal{F_{X^{\infty}}}, \mu, T_{\mathcal{X}}) $ is AMS. \\

\section* {Discrete Version of the RWMM}

We now initiate the study of a discrete time space version of the Random Waypoint Mobility Model. Consider a MANET with each mobile node in the set $\mathcal{V} = \{ v_1, v_2, ..., v_\mathcal{|V|} \}$ located in a discrete finite geographical area described by the set $\mathcal{S}$. The following are the random processes to exposit the discrete RWMM.

\subsection*{Waypoint Random Process Per Node}
From the geographical space $\mathcal{S}$, each mobile node $v \in \mathcal{V}$ selects an infinite tuple of waypoints $ \mathbf{w} = w_0,w_1,w_2,... $ randomly.  Let us denote the waypoint selection random process as $\mathbf{W} = \{W_k\}_{k=0}^\infty$ with the corresponding dynamical system as $  (\mathcal{W^{\infty}}, \mathcal{F_{W^{\infty}}}, \mu_\mathbf{w}, T_{\mathcal{W}}) $, and also $\mathcal{W = S}$. \\
\emph{RWMM Correlation :} In the classic RWM model, waypoint selection random process is IID, and in most cases uniformly distributed. So the stochastic process $  (\mathcal{W^{\infty}}, \mathcal{F_{W^{\infty}}}, \mu_\mathbf{w}, T_{\mathcal{W}}) $ is a Bernoulli Scheme\cite{WikiBer}.

\subsection*{Path Random Process Per Node}
\begin{figure}[h]
\centering
\includegraphics[width=30pc]{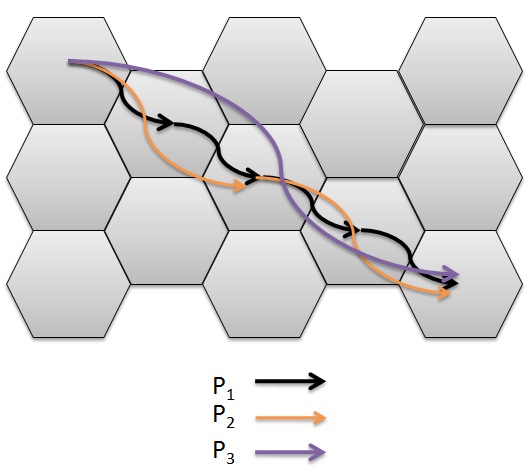}
\caption{Different paths corresponding to discrete time-space equivalent of different speeds}
\label{fig_dec}
\end{figure}
In the classic RWMM, whenever an arbitrary node selects a sequence of waypoints $\mathbf{w}$, then for each consecutive pair $(w_i,w_{i+1}), i \in \mathcal{Z^*}$, it also selects a speed uniformly distributed from $[min\_s,max\_s]$ and traverses the straight line path between $w_i$ and $w_{i+1}$. In discretized time and space, snapshot of the node's position per instance of time is taken during its trip between waypoints. This shall result in a random path with finite possibilities per waypoint pair $(w_i,w_{i+1})$ (figure 1). For each combination of waypoints $(w,w') \in \mathcal{W \times W}$ construct the set of all paths $\mathcal{P}_{w,w'}$ and take the union of all such sets so as to obtain all admissible paths $\mathcal{P}$:  $ \hspace*{15pt} \mathcal{P}_{w,w'} = \{ p_1, p_2, p_3, ..., p_{|\mathcal{P}_{w,w'}|} \} $, 
$ \hspace*{15pt} \mathcal{P} = \bigcup\limits_{(w,\acute{w}) \in \mathcal{W \times W}} \mathcal{P}_{w,\acute{w}}$.
In order to describe the stochastic process $\mathbf{P} =\{P_k\}_{k=0}^\infty$, noting that $\mathbf{P}$ is conditioned on $\mathbf{W}$, we first define the set of permitted path sequences $\mathcal{P}_\mathbf{w}^\infty\subset \mathcal{P}^\infty$ given $\mathbf{w}$ as 
$\mathcal{P}_\mathbf{w}^\infty = \{ \mathbf{p} \in \mathcal{P}^\infty : p_i \in \mathcal{P}_{w_i, w_{i+1}}, \forall i \in \mathcal{Z^*}\}$. Here again let $\mathcal{F_{P^\infty}}$ be the $\sigma$-algebra generated from $p_m^n \in \mathcal{P^\infty}$. Defining a collection of probability measures $\nu_\mathbf{wp} = \{ \nu_\mathbf{w} : \mathbf{w} \in \mathcal{W^\infty}, \nu_\mathbf{w}(\mathcal{P}_\mathbf{w}^\infty) = 1 \} $ results in the channel \cite{Ref14} 
$ (\mathcal{W}, \nu_\mathbf{wp}, \mathcal{P}) $. \\
\emph{Definition 5 (Stationary Channel)}\cite{Main}: If $ \forall \mathbf{w} \in \mathcal{W^\infty}, \forall A \in \mathcal{F_{P^\infty}}$, $ \hspace*{6pt} \nu_{T_\mathcal{W}\mathbf{w}}(A) = \nu_\mathbf{w}(T_\mathcal{P}^{-1}A) $ 
, the channel $ (\mathcal{W}, \nu_\mathbf{wp}, \mathcal{P}) $ is said to be $(T_\mathcal{W}, T_\mathcal{P})$ stationary.
\\\emph{RWMM Correlation:} Considering elementary events $[p_0^{n-1}] \in \mathcal{F_{P^\infty}}$: \\
\begin {displaymath}
\nu_\mathbf{w}([p_0^{n-1}]) = \prod_{i=0}^{n-1}  \frac {1} {|\mathcal{P}_{w_i,w_{i+1}}|} \hspace*{10pt}  p_i \in \mathcal{P}_{w_i,w_{i+1}}, 0 \le i \le n-1 \footnote[1]{Product upper limit error in \cite{Main}}
\end {displaymath}
And $\nu_\mathbf{w}$ is zero otherwise. To prove stationarity, see that on the transformed $T_\mathcal{W}\mathbf{w}$ the non-zero probability for  $[p_0^{n-1}]$ is given by: \\
\begin {displaymath}
\nu_{T_\mathcal{W}\mathbf{w}}([p_0^{n-1}]) = \prod_{i=0}^{n-1} \frac {1} {|\mathcal{P}_{w_{i+1},w_{i+2}}|} \hspace*{10pt}  p_i \in \mathcal{P}_{w_{i+1},w_{i+2}}, 0 \le i \le n-1 
\end {displaymath}
And if $T_\mathcal{P}^{-1}[p_0^{n-1}] = [\bar{p}_1^n]$ with $\bar{p}_{m+1} = p_m, 0 \le m < n$ : 
\begin {displaymath}
\nu_\mathbf{w}(T_\mathcal{P}^{-1}[p_0^{n-1}]) = \nu_\mathbf{w}([\bar{p}_1^n])= \prod_{i=0}^{n-1} \frac {1} {|\mathcal{P}_{w_{i+1},w_{i+2}}|} \hspace*{10pt}  p_i \in \mathcal{P}_{w_{i+1},w_{i+2}}, 0 \le i \le n-1 
\end {displaymath}
The last two equations are equal which proves that the channel is stationary.

Next it is proved that the channel is output mixing and consequently ergodic. A channel is said to be output mixing\cite{Main} if, 
$ \forall A,B \in \mathcal{F_{P^\infty}}\footnote[2]{Incorrect sigma field in \cite{Main}}, \forall \mathbf{w} \in \mathcal{W^\infty}$:
\begin {displaymath}
\lim_{n \to \infty} \left| \nu_\mathbf{w} \left(T_\mathcal{P}^{-n}A\bigcap B \right) - \nu_\mathbf{w} (T_\mathcal{P}^{-n}A) \nu_\mathbf{w} (B)  \right| = 0
\end {displaymath}
The elementary events in case of the general RWMM are decoupled for $\tau \geq b$ for $[p_0^{a-1}],[p_0^{b-1}]  \in \mathcal{F_{P^\infty}} $ in the following equation:
\begin {displaymath}
\nu_\mathbf{w} \left(T_\mathcal{P}^{-\tau}[p_0^{a-1}] \bigcap [p_0^{b-1}] \right) =
\nu_\mathbf{w} \left([\acute{p}_\tau^{\tau+a-1}] \bigcap [p_0^{b-1}] \right) = 
\end {displaymath}
\begin {displaymath} 
\left(\prod_{i = \tau}^{\tau + a - 2} \frac {1} {|\mathcal{P}_{w_{i+1},w_{i+2}}|} \right)
\left(\prod_{j = 0}^{b - 2} \frac {1} {|\mathcal{P}_{w_{j+1},w_{j+2}}|} \right) 
\begin {array} {c}
\acute{p}_i \in \mathcal{P}_{w_{i+1},w_{i+2}}, \tau \le i \le \tau+a-1 \\
p_j \in \mathcal{P}_{w_{j+1},w_{j+2}}, 0 \le j \le b-1 
\end {array}
\end {displaymath}
\begin {displaymath}
= \nu_\mathbf{w} \left(T_\mathcal{P}^{-\tau}[p_0^{a-1}]\right) \nu_\mathbf{w} \left([p_0^{b-1}] \right)
\end {displaymath}
Hence the channel is output mixing and ergodic \cite{Ref14}[Lemma 9.4.3].

 Finally we define a probability measure $\mu_\mathbf{p}$ conditioning it on the waypoint selection probability measure $\mu_\mathbf{w}$: $\mu_\mathbf{p}(A) = \sum\limits_{\mathbf{w'} \in \mathcal{W}^\infty}\mu_\mathbf{w}(\mathbf{w'}) \nu_\mathbf{w'}(A), \hspace*{5pt} \forall A \in \mathcal{F_{P^\infty}} $ .
Thus we result with $  (\mathcal{P^\infty}, \mathcal{F_{P^\infty}}, \mu_\mathbf{p}, T_\mathcal{P}) $ as the corresponding dynamical system for $\mathbf{P} = \{P_k\}_{k=0}^\infty$. \\

\subsection*{Location Random Process per node}
We define the time taken $t(i)$ to reach $w_i$ from $w_0$ as a function of the first $i$ paths $p_0, p_1, ..., p_{i-1}$. We assume that each path length $l(p)$ is a positive finite quantity. So $t(i) = \sum_{j=0}^{i-1}l(p_j), \hspace*{5pt}  i \ge 1$. Let the $i^{th}$ path $p_i$ take the form $s_{t(i)}, s_{t(i)+1}, s_{t(i)+2}, ..., s_{t(i+1)}$, with $s_j \in \mathcal{S}$ and $ s_{t(k)} = w_k $. Correlating with the given paths' sequence $\mathbf{p}$, we arrive at node location sequence $\mathbf{s} = s_0,s_1,...$. Thus we have the node location random process $\mathbf{S} = \{S_k\}_{k=0}^\infty$ given by the dynamical system  $  (\mathcal{S^\infty}, \mathcal{F_{S^\infty}}, \mu_\mathbf{s}, T_\mathcal{S}) $. 

\subsection*{Location Random Process for all nodes}
Consider the $\mathcal{|V|}$ tuple $ X_n = (S_{n,1},S_{n,2}, ... , S_{n,\mathcal{|V|}}) $, with $S_{i,j}$ denoting node $j$'s location at time $i$. This random variable's alphabet is $\mathcal{X} = \mathcal{S^{|V|}}$. Define 
$  (\mathcal{X^\infty}, \mathcal{F_{X^\infty}}, \mu_\mathbf{p}, T_\mathcal{X}) $ as the dynamical system for the random process $\{X_k\}_{k=0}^\infty$.

\section*{Main Result and its Proof}
\textsc {Theorem} \cite{Main}: Suppose the nodes $\mathcal{V} = \{ v_1, v_2, ..., v_\mathcal{|V|} \}$ move in agreement with the discrete RWMM already defined. Let $\mathbf{W}_v$ denote the waypoint selection random process for node $v$ and $\mathbf{P}_v$ be the corresponding path random process. Let $(\mathcal{W}_v, \nu_\mathbf{wp}, \mathcal{P}_v)$ denote the path and waypoint stochastic processes' connecting channel and let $\mathbf{X}$ denote the location random process for all nodes.  Then 
\begin {itemize}
\item If $\forall v \in \mathcal{V}$, $\mathbf{W}_v$ is AMS and the channel  $(\mathcal{W}_v, \nu_\mathbf{wp}, \mathcal{P}_v)$ is stationary, then $\mathbf{X}$ is AMS and stable.
\item If $\forall v \in \mathcal{V}$, $\mathbf{W}_v$ is ergodic and the channel  $(\mathcal{W}_v, \nu_\mathbf{wp}, \mathcal{P}_v)$ is ergodic, then $\mathbf{X}$ is ergodic.
\end {itemize}

\noindent \textsc {Proof Sketch} \cite{Main}:\\
Dropping the redundant subscript $v$ henceforth. \\

\emph{ Lemma A: If $\mathbf{W}$ is AMS and ergodic and $(\mathcal{W}, \nu_\mathbf{wp}, \mathcal{P})$ is stationary and ergodic, then $\mathbf{P}$ is AMS and ergodic.} \\
\emph{Proof:} \cite{Ref14}[Lemmas 9.3.1, 9.3.3] prove this lemma directly as the AMS and ergodic waypoint random process and the path random process are connected by a stationary, ergodic channel. \\

\emph{ Lemma B: If $\mathbf{P}$ is ergodic  then $\mathbf{S}$ is ergodic.} \\
\emph{Proof:} Given $  (\mathcal{P^\infty}, \mathcal{F_{P^\infty}}, \mu_\mathbf{p}, T_\mathcal{P}) $ is AMS. For all $p \in \mathcal{P}$ let $l(p)$ denote path length, let $L = \max\{ l(p) : \mathbf{p} \in \mathcal{P} \}$ and let $f:\mathcal{P} \to \bigcup_{i=1}^L\mathcal{S}^i$ be the breakdown of a path to its corresponding geographic cells - $f(p) = s_0,s_1,s_2,...,s_{l(p)-1}$. So 
$\mathbf{S} = \{S_k\}_{k=0}^\infty = f(P_0),f(P_1),f(P_2),... = S_0,S_1,...,S_{l(P_0)},...,S_{l(P_0)+l(P_1)},...$. \\
For ease of working, define the encoder $F:\mathcal{P}^\infty \to \mathcal{S}^\infty$ as $\mathbf{s} = F(\mathbf{p}) = f(p_0),f(p_1),f(p_2),...$. Now $\forall A \in \mathcal{F_{S^\infty}}, \mu_\mathbf{s}(A) = \mu_p(F^{-1}A)$. Here the mapping $F$ is many to one.\\
Next it is described a pseudo-inverse $G^{-1}: \hat{\mathcal{S}}^\infty \to \hat{\mathcal{P}}^\infty$ as $G^{-1}\mathbf{s} = \mathbf{p_s}$ where $\hat{\mathcal{P}}^\infty \subseteq \mathcal{P}^\infty$ according to \cite{Ref9}[Theorem 1] having full measure $\mu_\mathbf{p}( \hat{\mathcal{P}}^\infty) = 1$ such that every bounded measurable function on this set converges;  $\hat{\mathcal{S}}^\infty$ is the induced range of $F$ on $\hat{\mathcal{P}}^\infty$ and $\mathbf{p_s}$ is a representative from the partition of $\hat{\mathcal{P}}^\infty$ induced by $\mathbf{s} \in \hat{\mathcal{S}}^\infty$. \\ 
Define the length of the first $n$ paths in $\mathbf{p} \in \mathcal{P}^\infty$ as $\gamma_\mathbf{p}(n) = \sum_{i=0}^{n-1}l(p_i)\footnote[1]{Wrong limit in \cite{Main}}$. Then the variable length shift $T_\mathcal{S^*}:  \mathcal{S}^\infty \to  \mathcal{S}^\infty$ is given by $ T_\mathcal{S^*}^n\mathbf{s} = T_\mathcal{S}^{\Gamma_n(\mathbf{s})}\mathbf{s}$ where \\
\begin {displaymath}
\Gamma_n(\mathbf{s}) =
\begin {array} {c} 
\gamma_{G^{-1}(\mathbf{s})}(n), \hspace*{10pt} \mathbf{s} \in \hat{\mathcal{S^*}} \\
1, \hspace*{50pt} \mathbf{s} \notin \hat{\mathcal{S^*}}
\end {array}
\end {displaymath}
Eventually it is proved that $\lim_{n \to \infty} \frac {1} {n} \sum_{k=0}^{n-1} h(T_\mathcal{S^*}\mathbf{s})$ exists $\forall \mathbf{s} \in \hat\mathcal{S^*}$ and for all bounded measurable $h$. Thus $  (\mathcal{S^\infty}, \mathcal{F_{S^\infty}}, \mu_\mathbf{s}, T_\mathcal{S^*}) $ is AMS. Note that one $T_\mathcal{S^*}$ shift is equivalent to one path shift. \\
\emph{Sublemma: If $  (\mathcal{S^\infty}, \mathcal{F_{S^\infty}}, \mu_\mathbf{s}, T_\mathcal{S^*}) $ is AMS with stationary mean $\bar\mu_\mathbf{s}^*$ then $  (\mathcal{S^\infty}, \mathcal{F_{S^\infty}}, \mu_\mathbf{s}, T_\mathcal{S}) $ is AMS where $ T_\mathcal{S^*}\mathbf{s} = T_\mathcal{S}^{\gamma(\mathbf{s})}\mathbf{s} $  and $ 1 \le \gamma(\mathbf{s}) \le L \hspace*{3pt}$. \footnote[2]{Typographical error in \cite{Main} for $ T_\mathcal{S^*}$}
} \\
Define a new measure (inspired from \cite{Ref9}[Ex.6]):\\
\begin {displaymath}
\bar\mu_\mathbf{s}(A) = \frac {1} {E_{\bar\mu_\mathbf{s}}[\gamma(\mathbf{s})]} \sum_{k=1}^L\sum_{i=0}^{k-1} \bar\mu_\mathbf{s}^*(T_\mathcal{S}^{-i}A \cap \Delta_k^{-1}) 
\end {displaymath}
Here $\Delta_k^{-1} = \{ \mathbf{s} \in \mathcal{S^\infty} : \gamma(\mathbf{s}) = k\}$, and $\{\Delta_k^{-1}\}_{k=1}^L$ is a partition of $\mathcal{S^\infty}$ \cite{Ref9}. Thus we have 
$T_\mathcal{S^*}^{-1}A = \cup_{k=1}^L(T_\mathcal{S}^{-k}A \cap \Delta_k^{-1})$. We also have 
\begin {displaymath}
\bar\mu_\mathbf{s}^*(T_\mathcal{S^*}^{-1}A) = \bar\mu_\mathbf{s}^*(A) = \sum_{k=1}^L\bar\mu_\mathbf{s}^*(A \cap \Delta_k^{-1}) = \sum_{k=1}^L\bar\mu_\mathbf{s}^*(T_\mathcal{S}^{-k}A \cap \Delta_k^{-1})
\end {displaymath}
The first two terms are equal by virtue of transformation invariance of $\bar\mu_\mathbf{s}^*$. The next two terms are equal by virtue of intersection distribution of $A$ on $\Delta_k^{-1}$. The first and the last term are equal from the immediately preceding correlation between $T_\mathcal{S^*}$ and $T_\mathcal{S}$. Substituting $T_\mathcal{S}^{-1}A$ for $A$ in the definition of $\bar\mu_\mathbf{s}$ and using the above equation we arrive at the $T_\mathcal{S}$ invariance of $\bar\mu_\mathbf{s}$.
Further it is shown that $\bar\mu_\mathbf{s}$ asymptotically dominates $\bar\mu_\mathbf{s}^*$ under $T_\mathcal{S}$ which when taken with the $T_\mathcal{S}$ invariance of $\bar\mu_\mathbf{s}$ and \cite{Ref9}[Theorem 2] proves that $\bar\mu_\mathbf{s}^*$ is AMS w.r.t. $T_\mathcal{S}$    . \\
Next it is proved that if $\bar\mu_\mathbf{s}(A) = 0$ and $T_\mathcal{S}^{-1}A = A$ then $\mu_\mathbf{s}(A) = 0$. This in conjunction with \cite{Ref15}[Theorem 2.2] proves that $\mu_\mathbf{s}$ is AMS w.r.t. $T_\mathcal{S}$. \\
Thus from the sublemma $  (\mathcal{S^\infty}, \mathcal{F_{S^\infty}}, \mu_\mathbf{s}, T_\mathcal{S}) $ is AMS which completes the proof. \\

\emph{ Lemma C: If $\mathbf{P}$ is ergodic, then $\mathbf{S}$ is ergodic } \\
\emph {Proof: } Let $A \in \mathcal{F_{S^\infty}}$ be $T_\mathcal{S}$ invariant and let $\mathbf{s} = F(\mathbf{p}) $ be an arbitrary member of $A$. Now \\ 
$F(\mathbf{p}) \in A \Leftrightarrow T_\mathcal{S}^nF(\mathbf{p}) \in A \hspace*{3pt}$ 
$ \Rightarrow F(\mathbf{p}) \in A \Leftrightarrow T_\mathcal{S}^{l(p_0)}F(\mathbf{p}) \in A \hspace*{3pt}$  
$ \Rightarrow F(\mathbf{p}) \in A \Leftrightarrow F(T_\mathcal{P}\mathbf{p}) \in A$  \\
Taking $F^{-1}$ on both sides (as the equation holds for all $\mathbf{s}$ and all $\mathbf{p}$ associated with each $\mathbf{s}$ )\\
$ \mathbf{p} \in F^{-1}A \Leftrightarrow T_\mathcal{P}\mathbf{p} \in  F^{-1}A $ . Hence $F^{-1}A$ is $T_\mathcal{P}$ invariant. By the premise of the lemma, $\mu_\mathbf{p}(F^{-1}A) = 0$ or $1$. Hence $\mu_\mathbf{s}(A) = 0$ or $1$. Thus by definition, $ (\mathcal{S^\infty}, \mathcal{F_{S^\infty}}, \mu_\mathbf{s}, T_\mathcal{S}) $ is ergodic.

\section*{Simulation}
We have simulated a basic packet exchange in a MANET using NS2 with the node movement generated according to the general continuous RWMM. The traffic consisted of constant bitrate UDP packets with IEEE 802.11 protocol at the MAC layer. The exchanges resulted in bursty traffic. The plot for the number of bytes received as a function of time for a particular node is given in figure 2.
\begin{figure}[h]
\centering
\includegraphics[width=30pc]{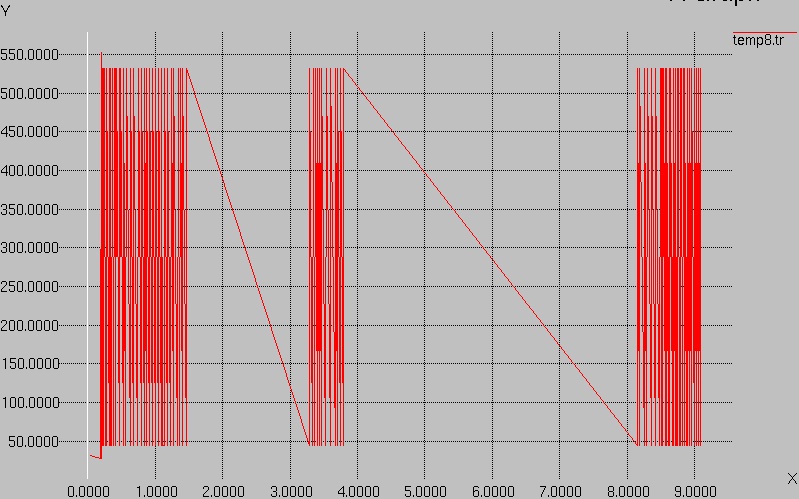}
\caption{Number of bytes received as a function of time for a particular node}
\end{figure}

\section*{Conclusion}
In this paper we have successfully demonstrated that the discrete general RWMM is AMS and hence stable. Thus simulations with RWMM as the underlying node movement generation algorithm tend to be reliable. The stability preserving protocols allow higher layers of the protocol stack to propagate this stability hence permitting reliability of simulations at higher levels also. Moreover we have simulated the continuous version of the RWMM with the intent of seeing the local non-stationary properties (which is highlighted by the bursty traffic).

\end {document}